\documentclass[aps,prd,nofootinbib,superscriptaddress,twocolumn]{revtex4-1}

\usepackage{amsmath}
\usepackage{amssymb}
\usepackage{bbold}
\usepackage{epsfig}
\usepackage{nicefrac}
\usepackage{graphicx}
\usepackage{color}
\usepackage{epstopdf}
\usepackage{tabularx}
\usepackage{dcolumn}  
\newcolumntype{L}[1]{>{\raggedright\arraybackslash}p{#1}}
\newcolumntype{C}[1]{>{\centering\arraybackslash}p{#1}}
\newcolumntype{R}[1]{>{\raggedleft\arraybackslash}p{#1}}

\usepackage[colorlinks=true, pdfstartview=FitV, linkcolor=blue, citecolor= red,urlcolor=blue]{hyperref}
\definecolor{darkgreen}{rgb}{0,0.5,0}
\definecolor{purple}{rgb}{0.5,0,0.5}
\definecolor{nblue}{rgb}{0.0,0.0,0.50}
\definecolor{scarlet}{rgb}{1.0,0.2,0}

\begin{document}

\title{Pion fragmentation functions from a quark-jet model in a functional approach}

\author{Roberto Correa da Silveira}
\affiliation{Labor\'at\'orio de F\'isica Te\'orica e Computacional , Universidade Cidade de S\~ao Paulo, Rua Galv\~ao Bueno 868, S\~ao Paulo, S\~ao Paulo 01506-000, Brazil}
\affiliation{Instituto de F\'{\i}sica Te\'orica, Universidade Estadual Paulista, Rua Dr.~Bento Teobaldo Ferraz 271, 01140-070 S\~ao Paulo, S\~ao Paulo, Brazil}

\author{Fernando E. Serna}
\affiliation{Departamento de F\'isica, Universidad de Sucre, Carrera 28 No.~5-267, Barrio Puerta Roja, Sincelejo 700001, Colombia}

\author{Bruno El-Bennich}
\affiliation{Instituto de F\'{\i}sica Te\'orica, Universidade Estadual Paulista, Rua Dr.~Bento Teobaldo Ferraz 271, 01140-070 S\~ao Paulo, S\~ao Paulo, Brazil}
\affiliation{Departamento de F\'isica, Universidade Federal de S\~ao Paulo, Rua S\~ao Nicolau 210, Diadema, 09913-030 S\~ao Paulo, Brazil}


\begin{abstract}
The elementary fragmentation function that describes the process $q\to \pi$ is predicted applying crossing and charge symmetry to the cut diagram 
of the pion valence quark distribution function. This elementary probability distribution defines the ladder-kernel of a quark jet fragmentation equation which is
solved self-consistently to obtain the full pion fragmentation function. The hadronization into a pion employs the complete Poincar\'e invariant Bethe-Salpeter
wave function, though the overwhelming contribution to the fragmentation function is due the leading Bethe-Salpeter amplitude. Compared to a Nambu--Jona-Lasinio 
model prediction, the fragmentation function we obtain is enhanced in the range $z \lesssim 0.8$ but otherwise in good  qualitative agreement. The full pion fragmentation 
function is overall greater than the elementary fragmentation function below $z\lesssim 0.6$.
\end{abstract}

\date{\today}
\maketitle


\section{Introduction}
\label{intro}

Many high-energy scattering processes can be treated perturbatively in Quantum Chromodynamics (QCD) within the framework of factorization theorems. 
The latter allow for a separation of the perturbatively calculable part of a given cross section from the nonperturbative matrix elements at a given energy scale, 
where the latter are frequently codified in parton distribution functions. They also are fundamental tools  in understanding the internal 
structure of protons and other hadrons. 

Likewise, quark jet fragmentation functions describe how the energy and momentum are distributed amongst highly energetic hadrons that form a jet with nearly parallel 
longitudinal momenta and negligible transverse momenta~\cite{Field:1976ve}. Such jets are produced in high-energy collisions where reaction and decay processes 
produce gluon and quark partons that fragment into a cascade of hadrons once they escape the interaction region. Besides being of fundamental interest in their own 
right, their knowledge is essential for the extraction of the transversity quark distribution functions~\cite{Barone:2001sp,Matevosyan:2011vj}. At the heart of quark jet 
fragmentations functions are elementary fragmentation functions of quarks and gluons~\cite{Metz:2016swz}, which describe the probably for such a parton to fragment 
into a single hadron with a given fraction of its light-front momentum. 

In here, we obtain the elementary $q\to \pi$ fragmentation function from the corresponding cut diagram~\cite{Jaffe:1983hp,Bentz:1999gx,Ito:2009zc} using a functional 
approach to QCD modeling. In doing so, we solve the Dyson-Schwinger equation (DSE) for a light quark, while the solution of the Bethe-Salpeter equation (BSE) in the 
pseudoscalar channel yields the pion wave function. The $q\to \pi$ fragmentation function for a physical light-front momentum fraction $z$ and the parton distribution function 
of a quark in a pion for unphysical $x>1$ obey a Drell-Levy-Yan (DLY) relation~\cite{Drell:1969jm,Drell:1969wd,Drell:1969wb} due to charge conjugation and crossing symmetry.
However, in a single step fragmentation process, such as $q\to \pi$, momentum and isospin  sum rules cannot be satisfied~\cite{Ito:2009zc} and one must consider the 
possibility that the fragmenting quark produces a cascade of mesons. The infinite fragmentation tower can in principle involve any hadrons, though we here limit ourselves 
to the hadronization into a cascade of pions described by jet functions,  much alike the original quark jet-model of Fields and Feynman~\cite{Field:1976ve}. 
Our approach satisfies momentum and isospin sum rules and represents a systematic extension of a Nambu--Jona-Lasinio (NJL) model calculation~\cite{Ito:2009zc} 
where the Poincar\'e invariant Bethe-Salpeter amplitude is not point-like but possesses a spatial extension and the quark masses are momentum dependent.  

In Section~\ref{sec2} we present the theoretical framework, namely the DSE and BSE that describe the quark and the bound state wave function of the pion, respectively. 
The elementary fragmentation function we obtain is discussed in Section~\ref{sec3} and our prediction is compared to that of a NJL model~\cite{Ito:2009zc}.
We emphasize that while the latter is about 10--12\% suppressed compared with the fragmentation function obtained in the DSE-BSE approach, the qualitative 
behavior of both functions is very similar. This is chiefly due to the fact that the fragmentation function is dominated by the contribution of the leading Bethe-Salpeter 
amplitude, which mirrors the NJL calculation where the wave function is a single, point-like Lorentz covariant. We then discuss the quark-jet fragmentation equations
from which we obtain full fragmentation function that describes the infinite tower of quark hadronization into pions of different isospin. We wrap up with concluding 
remarks  in Section~\ref{sec4}.


\section{From quark to pion: gap and bound-state equations}
\label{sec2}


\subsection{Dressed quark propagators}

In a nonperturbative approach to the quark fragmentation function, the fully dressed quark propagator is a crucial ingredient. We obtain this propagator from the 
corresponding DSE of a dressed quark,

 \begin{align} 
   S^{-1} (p)  =  & \ Z_2 \! \left (i\, \gamma \cdot  p + m_\mathrm{bm} \right )  \nonumber   \\
                    +  &\,  Z_1\,  g^2 \!\! \int^\Lambda \!\! \frac{d^4k}{(2\pi)^4}  \, D^{ab}_{\mu\nu} (q) \frac{\lambda^a}{2} \gamma_\mu \,S(k) \, \Gamma^b_{\nu}  (k,p) \, ,
\label{QuarkDSE}
\end{align}
whose solutions are Euclidean Schwinger functions~\cite{Bashir:2012fs,Cloet:2013jya}. In Eq.~\eqref{QuarkDSE}, $m^\textrm{bm}$ is the bare current-quark 
mass, $Z_1(\mu,\Lambda)$ and $Z_2(\mu,\Lambda)$ are the vertex and wave-function renormalization constants at the renormalization scale $\mu$, respectively. 
The integral describes  the self-energy of the quark and involves the dressed-quark propagator $S(k)$, the dressed-gluon propagator $D_{\mu\nu}(q)$ with $q=k-p$
and the quark-gluon vertex, $\Gamma^a_\mu (k,p) = \frac{1}{2}\,\lambda^a  \Gamma_\mu (k,p)$~\cite{Albino:2018ncl,Albino:2021rvj,El-Bennich:2022obe,Lessa:2022wqc}, 
where  $\lambda^a$ are SU(3) color matrices. The regularization scale  $\Lambda \gg \mu$ can be taken to infinity. For a given flavor, the general covariant 
form of the dressed propagator is,
\begin{equation}
    S (p)  =   - i \sigma_V ( p^2 )  \gamma \cdot p + \sigma_S ( p^2 ) =  \frac{ Z (p^2 )}{ i \gamma \cdot p + M ( p^2 )}\, ,
\label{DEsol}                        
\end{equation}
where $Z (p^2)$ is the wave renormalization function and $M (p^2)$ is the mass function of the quark, and the functions $\sigma_S( p^2 )$ 
and $\sigma_V ( p^2 )$ can be related to them. In a subtractive renormalization scheme one imposes the conditions,
\begin{align}
   Z( \mu^2)  & \equiv \, 1 \  ,
\label{EQ:Amu_ren}   \\
   S^{-1} ( \mu^2 ) & \equiv  \,  i \gamma\cdot p \ + m (\mu ) \ ,
 \label{massmu_ren}
\end{align}
where $m (\mu )$ is the renormalized current-quark mass related to the bare mass by,
\begin{equation}
\label{mzeta} 
   Z_4 (\mu,\Lambda )\, m (\mu)  = Z_2 (\mu,\Lambda ) \, m_{\rm bm} (\Lambda) \  ,
\end{equation}
and $Z_4 (\mu,\Lambda )$ is the mass-renormalization constant in the QCD Lagrangian.

The leading rainbow truncation of the DSE~\eqref{QuarkDSE} and the ladder truncation of the corresponding BSE kernel has proven to be a robust approximation 
that preserves the axialvector Ward-Green-Takahashi identity and allows for the description of light ground-state mesons in the isospin-nonzero pseudoscalar 
and vector channels. In essence, the fully dressed quark-gluon vertex is reduced to the perturbative vertex,  $\Gamma_\nu \to Z_2 \gamma_\nu$. Therefore,
the DSE kernel becomes~\cite{Serna:2018dwk},
\begin{equation}
   Z_1 g^2  D_{\mu\nu} (q)\, \Gamma_\nu (k, p) =  Z_2^2\,  \mathcal{G} (q^2) D_{\mu\nu}^\mathrm{free} (q) \frac{\lambda^a }{2} \gamma_\nu \, .
\label{RLtrunc}
\end{equation}
This truncation is characterized by an Abelianized Ward identity and, as in QED, implies $Z_1 = Z_2$~\cite{Bashir:2012fs} due to the omission of the three-gluon 
interaction in $\Gamma_{\mu}(k, p)$.  The additional factor $Z_2$ in Eq.~\eqref{RLtrunc} ensures multiplicative renormalizability of the DSE and consequently 
$M(p^2)$ is a renormalization-point invariant quantity~\cite{Bloch:2002eq}.

The free-gluon propagator is transverse in Landau gauge,
\begin{equation}
    D_{\mu\nu}^\mathrm{free} (q) =  \delta^{a b}\left(\delta_{\mu \nu}-\frac{q_{\mu} q_{\nu}}{q^{2}}\right ) \! \frac{1}{q^2} \ ,
 \label{freegluon}   
\end{equation}    
and the dressing function $\mathcal{G} (q^2)$ is composed of two pieces,
\begin{equation}
    \frac{\mathcal{G} (q^2)}{q^2}  = \,  \mathcal{G}_\mathrm{IR} (q^2) +  4\pi \tilde\alpha_\mathrm{PT} (q^2)  \ ,
\label{IR+UV}    
\end{equation}
where we deliberately absorb a factor $1/q^2$ of the gluon propagator~\eqref{freegluon}. The first term in Eq.~\eqref{IR+UV} dominates in the infrared domain, 
$q^2 < \Lambda^2_\mathrm{QCD}$, and is suppressed at large momenta, while the second term describes a bounded and monotonically decreasing continuation 
of the perturbative coupling that dominates at large momenta: $\mathcal{G}_{\mathrm{IR}} (q^2) \ll \tilde{\alpha}_{\mathrm{PT}} (q^2) \, \forall \, q^2 \gtrsim 2 \
\mathrm{GeV}^2$. In here, we use the infrared-finite dressing model of Ref.~\cite{Qin:2011dd} which reads,
\begin{align}
    \mathcal{G}^\mathrm{IR} (q^2)  & =   \frac{8\pi^2}{\omega^4}  D\, e^{-q^2/\omega^2}    
\label{G-IR}      \\
  4\pi \tilde\alpha_\mathrm{PT}(q^2) & =  \frac{8\pi^2  \gamma_m\, \mathcal{E}(q^2)}{ \ln \left [  \tau +
                       \left (1 + q^2/\Lambda^2_\textrm{\tiny QCD} \right )^{\!2} \right ] }  \ , 
\label{G-PT}                         
\end{align}
where $\gamma_m=12/(33-2N_f)$ is the anomalous mass dimension and $N_f $ is the active flavor number, $\Lambda_\textrm{\tiny QCD}=0.234$~GeV, 
$\tau=e^2-1$, $\mathcal{E}(q^2)=[1-\exp(-q^2/4m^2_t)]/q^2$, $m_t=0.5$~GeV, $\omega= 0.5\,$GeV  and $\kappa=(\omega D)^{1/3} = 0.8\,$GeV. 

In solving the DSE~\eqref{QuarkDSE}, we determine the renormalization constants $Z_2$ and $Z_4$ with the conditions $A(2\,\mathrm{GeV}) =1$ and $m(2\,\mathrm{GeV}) 
\equiv B(2\, \mathrm{GeV}) = 0.018$~GeV~\cite{Serna:2020txe}.


\subsection{Pion wave function}

The pions produced in the fragmentation of a quark are described by a Poincar\'e-covariant Bethe-Salpeter amplitude (BSA), $\Gamma_\pi (k,p)$. The latter is the solution of 
the homogeneous BSE for pseudoscalar mesons,
\begin{equation}
  \Gamma_\pi (k,p) =  \int^\Lambda \!\!  \frac{d^4q}{(2\pi)^4} \, K (k,q,p) \, \chi_\pi  (q,p)  \, ,  
\label{BSE-pseudo}
\end{equation} 
where $k$ is the relative quark-antiquark momentum, $p$ is the on-shell pion momentum, $p^2= -m_\pi^2$, and $K(k,q,p)$ is the fully amputated scattering kernel. 
The Bethe-Salpeter wave function of the pion, $\chi_\pi (k,p)$ is defined by attaching the quark propagators to the BSA,
\begin{equation}
   \chi_\pi (k,p)  \equiv  S (k_\eta)\, \Gamma_\pi (k,p) S (k_{\bar\eta}) \, ,
\end{equation} 
with the  momentum-partition parameters $\eta + \bar\eta = 1$ and the quark and antiquark momenta $k_{\eta} = k+\eta p$ and $k_{\bar\eta} = k-\bar\eta p$.
The ladder truncation compatible with the rainbow truncation of the DSE in Eq.~\eqref{RLtrunc} reads, 
\begin{equation}
    K (k,q,p)   = - Z_2^2 \,\frac{\mathcal{G} (l^2)}{l^2 }\, D_{\mu\nu}^\mathrm{free} (l)\, \frac{\lambda^a }{2} \gamma_\mu \frac{\lambda^a }{2} \gamma_\nu \ ,
  \label{RLkernel}    
\end{equation}
and $l=k-q$ is the gluon momentum exchanged infinitely many times between the quark and antiquark. 

The BSA is composed of Dirac matrices and the relative and total momenta consistent with the quantum numbers $J^{PC} = 0^{-+}$. Its most general 
Poincar\'e-covariant form is given by~\cite{Llewellyn-Smith:1969bcu},
\begin{align}
   \Gamma_\pi (k,p)  & =   \gamma_5    \Big [ i E_\pi (k,p) +  \gamma \cdot p\,  F_\pi (k,p)   \nonumber \\  
                                 + & \  \gamma \cdot k  \, k \cdot p \,   G_\pi (k,p)  +   \sigma_{\mu\nu} k_\mu p_\nu \, H_\pi (k,p) \Big ]  ,
 \label{PS-BSA}                                      
\end{align}     
in which $E_\pi, F_\pi, G_\pi, H_\pi$ are Lorentz-invariant scalar amplitudes. In the calculation of the elementary fragmentation function $d_{q}^{\pi}(z)$~\eqref{EQ:fragLF},
we employ the BSE solution of Ref.~\cite{Serna:2020txe} for the pion.


\section{Fragmentation functions}
\label{sec3} 


\subsection{Elementary quark fragmentation function}

Fragmentation functions describe the hadronization of highly energetic partons with nearly longitudinal momenta, though the transverse momentum dependence can 
be included~\cite{Matevosyan:2011vj}. The original parton model of fragmentation~\cite{Field:1976ve} is nowadays understood as the process of very energetic quarks 
and  gluons escaping the initial collision region and then fragmenting into jets of hadrons. 

\begin{figure}[t!] 
\centering
  \includegraphics[scale=0.26,angle=0]{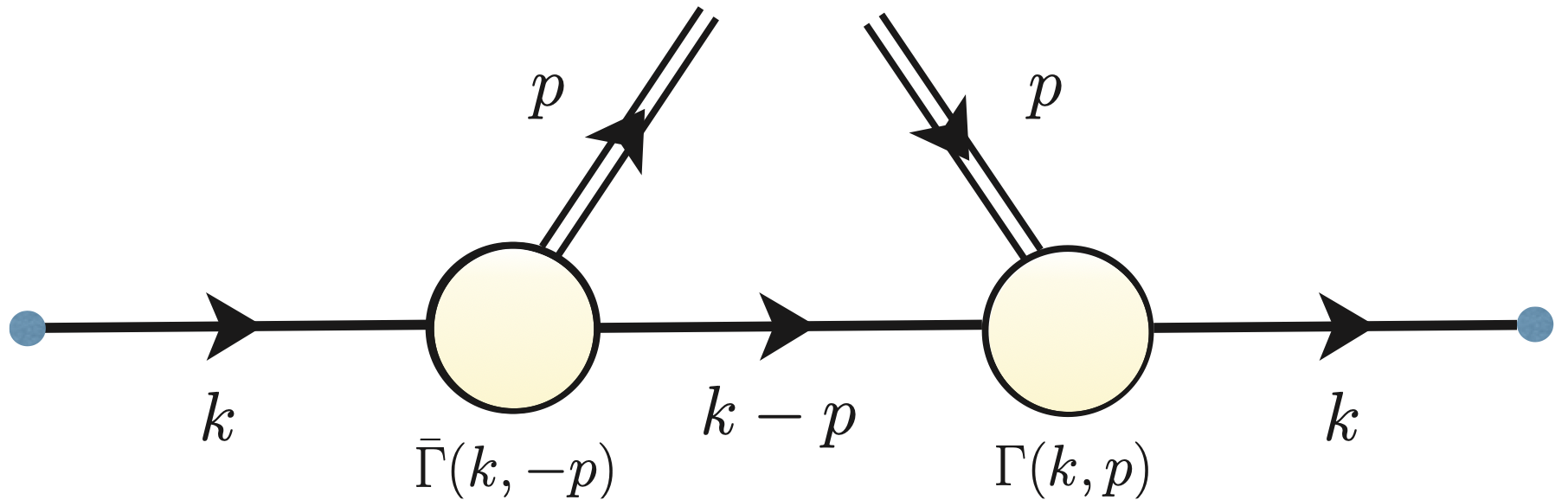} 
\caption{Cut diagram of the fragmentation function $d_{q}^{\pi} (z)$  in Eq.~\ref{EQ:fragLF}.  Shaded circle with outgoing/incoming double-solid line denote the pion
              in the fragmentation process, solid lines are quark propagators and the solid dots represent the $\gamma^+$ between the two quarks with momentum $k$.  }
\label{fragdiagram} 
\end{figure}

The elementary fragmentation function of the process $q \rightarrow q\pi$ for physical light-front momentum fraction $z=1/x <1$ can be related to the parton distribution 
function for \emph{unphysical} $x>1$ using crossing and charge-conjugation symmetry,
\begin{equation}
    d_q^\pi (z) =  \frac{z}{6}\, f_q^\pi (x) \, ,
\end{equation}
which is the DLY relation~\cite{Drell:1969jm,Drell:1969wd,Drell:1969wb}, and $x=k^+ / p^+$ is the fraction of light-front momentum transferred from a 
valence quark to a pion. This elementary fragmentation function represents a probability density and can be formulated with the cut diagram depicted in Fig.~\ref{fragdiagram} 
whose evaluation leads to the expression~\cite{Ito:2009zc}:
\begin{align} 
   d_{q}^{\pi} (z)  = & \, \frac{N_c   C_q^\pi z}{6}\! \int \! \frac{d^4 k}{(2\pi)^4 } \,   2\pi i\, \delta\Big (k^{+} - \tfrac{p^+}{z} \Big )  \delta\! \left ( \ell^2 + M^2 (\ell^2 ) \right )  
       \nonumber \\
    &   \times   \operatorname{Tr}_\text{D}  \Big [  S(k)\gamma^{+}S(k)\,\bar{\Gamma}_{\pi} \big (-k+\tfrac{p}{2}, - p \big )  
       \nonumber \\
    &  \times  \big ( -i \gamma \cdot \ell  + M (\ell^2 ) \big ) \, \Gamma_{\pi} \big (k-\tfrac{p}{2}, p \big )  \Big  ] . 
\label{EQ:fragLF}   
\end{align}
In Eq.~\eqref{EQ:fragLF}, $N_c=3$, $k$ and $\ell = k-p$ are the incoming and outgoing four-momenta of the quark, $S(k)$  is  the fully dressed quark 
propagator~\eqref{DEsol}, $\Gamma_\pi (k-\tfrac{p}{2},p)$ and $\bar \Gamma_\pi (-k+\tfrac{p}{2},-p)$ are the BSA~\eqref{PS-BSA}  and  the charge-conjugate 
BSA of the pion with total momentum $p$, respectively. $C_q^\pi = (1+\tau_q \tau_\pi )/2$ is an isospin factor. The light-front momenta are defined as $k^{+}=n\cdot k$, 
$p^{+}=n\cdot p$ and $\gamma^{+}=\gamma \cdot n$,  where $n$ is a light-like four-vector: $n^2=0$.

In order to calculate the fragmentation function~\eqref{EQ:fragLF} in light-front variables, a suitable algebraic representation of the BSA and quark propagators is 
indispensable\footnote{Alternatively, we can compute moments $\langle z^m\rangle = \int_0^1  z^m d_{q}^{\pi}(z) dz$ of the fragmentation function, which implies an 
integration over the Dirac function $\delta(k^+ - p^+ /z)$ in Eq.~\eqref{EQ:fragLF} followed by an integration over $d^4 k$ without resorting to light-front variables. 
With an adequate amount of moments the fragmentation function can then be reconstructed analogously to the case of Light Front Distribution Amplitudes~\cite{Serna:2020txe,Serna:2022yfp}. On the other hand, the reconstruction introduces  an uncertainty due to the fitting procedure we here eschew.}, 
as the dressing functions and scalar amplitudes are only available as numerical functions of their momentum squared and angles between the relative and total 
momentum of the meson~\cite{Serna:2020txe,Serna:2021xnr,Serna:2022yfp,daSilveira:2022pte,Serna:2024vpn}. As discussed in more detail in appendix~\ref{appendix1}, 
the generalized Nakanishi integral representation we use for the BSA and the complex-conjugate pole parametrization of the quark propagators depend on squared 
momenta and scalar products of momenta which we define in the following. 

We follow Ref.~\cite{Eichmann:2021vnj} and define light-front variables in Euclidean metric: 
\begin{equation}
   p^{ \pm} = -i p_4 \pm p_3  \, , \ \ 
   p_3 = \frac{p^+ - p^-}{2}\, ,    \ \
   p_4 = i\, \frac{p^+ +p^-}{2} \, ,  
\end{equation}
where $p_E = ( \boldsymbol p_T, p_3, p_4 )$. In the rest frame of the pion, $p = ( \boldsymbol 0, 0, i m_\pi )$, and therefore $p^\pm = m_\pi$. Evaluating the trace in 
Eq.~\eqref{EQ:fragLF} leads to scalar products which in light-front coordinates read, 
\begin{align}
   k_E \cdot p_E & =  \boldsymbol{k}_T \cdot \boldsymbol{p}_T -\frac{1}{2}  \left ( k^{-} p^{+}+k^{+} p^{-} \right )   \, ,   \label{LFkdotp} \\
   k_E^2   =  \boldsymbol{k}_T^{\,2}  & - k^+ k ^-   , \quad   p_E^2  = \boldsymbol{p}_T^{\,2}  - p^+ p ^-  = -m_\pi^2 \ ,  \label{LFk2}
\end{align}
while the four-momentum integral is:
\begin{equation}
    \int d^4k = \frac{1}{2} \int d\boldsymbol{k}_T\, dk^+ dk^- \, . 
\end{equation}
As discussed in Refs.~\cite{Collins:1981uw,Barone:2001sp}, Eq.~\eqref{EQ:fragLF} refers to the  frame where the produced meson has no transverse momentum, 
$\boldsymbol{p}_T = 0$, whereas the fragmenting quark is characterized by non-zero $\boldsymbol{k}_T$. In order to interpret this fragmentation function as a 
distribution of the pion in the quark, the quark has to be boosted to the frame where $\boldsymbol{k}_\perp = 0$. This can be achieved with a Lorentz transformation
and it follows that one may simply substitute,
\begin{equation}
    \boldsymbol{k}_T = -\frac{\boldsymbol{p}_{\perp}}{z} \, . 
\label{boost}    
\end{equation}
Note that we distinguish between both frames with the subscripts $T$ and $\perp$, that is $\boldsymbol{k}_T \neq 0$ but $\boldsymbol{k}_\perp =0$.

\begin{figure}[t!] 
\vspace*{-7mm}
\hspace*{-7mm}
\centering
  \includegraphics[scale=0.57,angle=0]{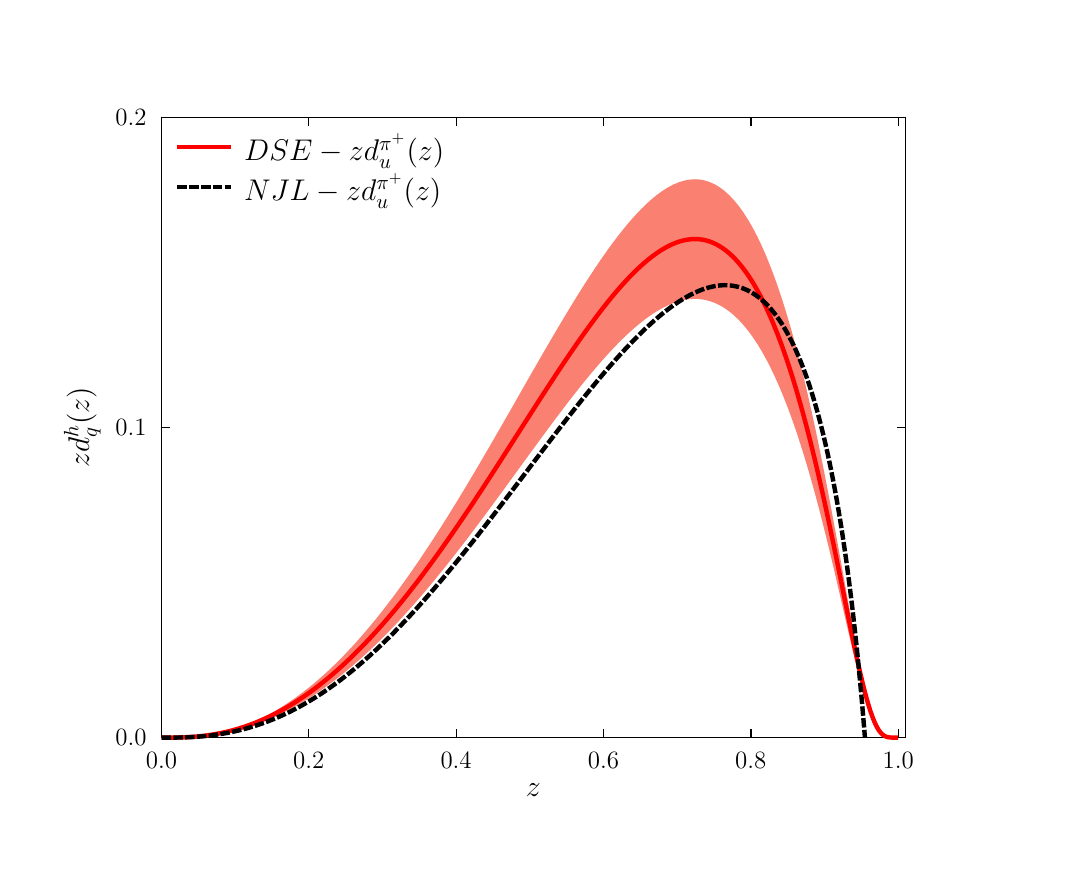} 
  \vspace*{-1.2cm}
\caption{The elementary $u \to \pi^+$ fragmentation function $d_u^{\pi^+}(z)$ derived from the expression in Eq.~\eqref{EQ:fragLF} compared with a prediction in the 
NJL model~\cite{Ito:2009zc}. The red-shaded error band stems from the systematic error in fitting the generalized Nakanishi integral representation~\eqref{NakanishiBSA}, 
as discussed in Ref.~\cite{Serna:2020txe}.}
\label{Dfragfig} 
\end{figure}

We recall that the quarks described by a running mass functions~\eqref{DEsol}  are confined and therefore there is no solution of the mass-shell equation
$ \ell^2 + M^{2}(\ell^2)$. On the other hand,  $\ell^2 = M^{2}(\ell^2)$, provides a working definition for a Euclidean-quark mass, and therefore a
realistic estimate of the quark's active constituent mass. We here use: $|\ell| = M_E= 0.41~\text{GeV}$~\cite{Serna:2018dwk}. 
With this, the second Dirac function in Eq~\eqref{EQ:fragLF}, using Eqs.~\eqref{LFkdotp}, \eqref{LFk2}, and \eqref{boost}  and $ \boldsymbol{p}_T  = \boldsymbol{0}$, 
becomes,
\begin{align}
   \delta \left ( \ell^2 + M^2 (\ell^2 )  \right ) & =   \nonumber \\
   \delta  \big ( \boldsymbol{p}_{\perp}^{\,2} / z^2 -k^+ & k^-  +  k^+p^- + k^- p^+ -m_\pi^2 + M_E^2 \big ) \, .
\label{deltafunc1}     
\end{align}
The first Dirac function in Eq~\eqref{EQ:fragLF} leads to $k^+ = p^+/z$ after integration over $k^+$ which modifies the second Dirac function to,
\begin{equation}
    \frac{z \delta\! \left ( k^- -  \dfrac{1}{(1-z) p^+} \left [\dfrac{  \boldsymbol{p}_{\perp}^2}{z} + (1-z) m_\pi^2 + z M_E^2 \right ] \right )}{p^+ (z-1)}  \, .
\end{equation}
Integrating over $k^-$, $k_E^2$ and $ k_E \cdot p_E$ become:
\begin{align}
    k_E \cdot p_E  & =  \frac{ \boldsymbol{p}_{\perp}^2 -m_\pi^2 \left ( z^2-1\right)+z^2 M_E^2}{2 z (z-1)} \, ,  \\
    k^2_E  & =  \frac{  \boldsymbol{p}_{\perp}^2 -m_\pi^2 (z-1)+z M_E^2}{z (z-1) } \, .
\end{align}
With this, the integral over $d\boldsymbol{k}_T = - d\boldsymbol{p}_\perp/z$ can be calculated numerically and normalized according to, 
\begin{equation}
   \sum_m \int_0^1 \hat {d}_q^m (z)\,  dz  = 1 \, ,  \quad m=\pi^+, \pi^0, \pi^-  ,
  \label{normsum} 
\end{equation}
which satisfies isospin and momentum sum rules. We have used the isospin notation $(\tau_u, \tau_d )=(1,-1)$ and $( \tau_{\pi^+}, \tau_{\pi^0}, \tau_{\pi^-} ) = (1,0,-1)$ 
in Eqs.~\eqref{EQ:fragLF} and \eqref{normsum}.

\begin{figure}[t!] 
\hspace*{-7mm}
  \includegraphics[scale=0.5,angle=0]{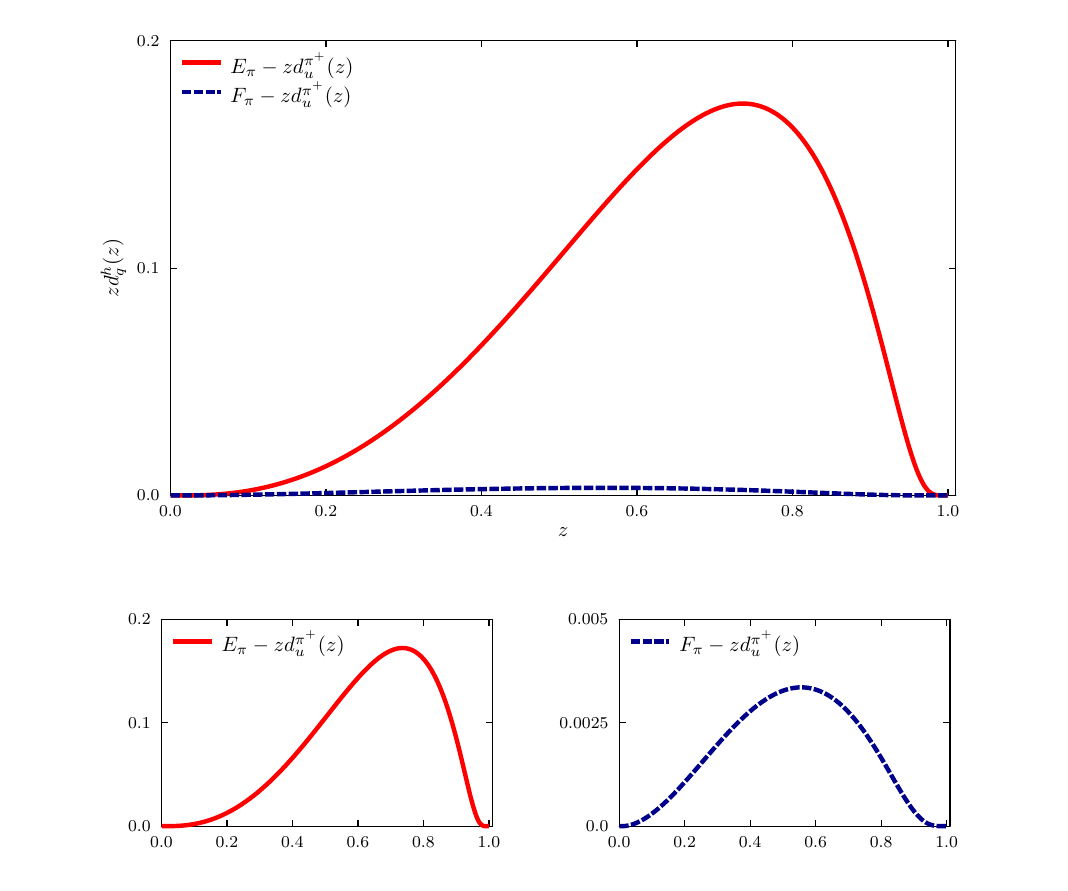} 
\caption{$E_\pi (k,p)$ and $F_\pi (k,p)$ contributions~\eqref{PS-BSA} to the $u \to \pi^+$ fragmentation function. The contributions of $G_\pi (k,p)$
and $H_\pi (k,p)$ are not plotted, as they are negligible. }
\label{DfragEandF-fig} 
\end{figure}

The  elementary fragmentation function $d_{q}^{\pi}(z)$ is compared to that obtained with a NJL model~\cite{Ito:2009zc} in Fig.~\ref{Dfragfig}. Both predictions are 
qualitatively in good agreement, though the fragmentation function obtained in the present work is clearly enhanced up to $z\simeq 0.8$ and slightly softer for larger 
$z$. In order to understand the similarity of both predictions, we plot the contributions of $E_\pi (k,p)$ and $F_\pi (k,p)$~\eqref{PS-BSA}  to $d_u^{\pi^+}$ in 
Fig.~\ref{DfragEandF-fig}.  Clearly, the fragmentation function is dominated by the leading amplitude, not unlike the single covariant though point-like pion 
wave function employed in the NJL calculation~\cite{Ito:2009zc}. 

Since we use a gluon-model interaction in Eq.~\eqref{IR+UV},  the renormalization scale of the quark propagators $S_f(k)$ is not that of a  given scheme in perturbation 
theory. Following Ref.~\cite{Ito:2009zc,Shi:2018zqd}, we also compute in the present approach the pion's PDF. We then match the first moment of the 
valence quark extracted from a $\pi N$ Drell-Yan analysis, $2 \langle x \rangle_v = 0.47(2)$~\cite{Sutton:1991ay,Gluck:1999xe}, by using next-to-leading order 
Dokshitzer-Gribov-Lipatov-Altarelli-Parisi (DGLAP) evolution~\cite{Botje:2010ay} including strange- and charm-quark mass thresholds from the experimental scale, 
 $Q = 2\,$GeV, to our model scale.  For the latter we find $Q_0 = 0.63\;$GeV.


\begin{figure}[t!] 
\vspace*{-8mm}
\hspace*{-4mm}
  \includegraphics[scale=0.57,angle=0]{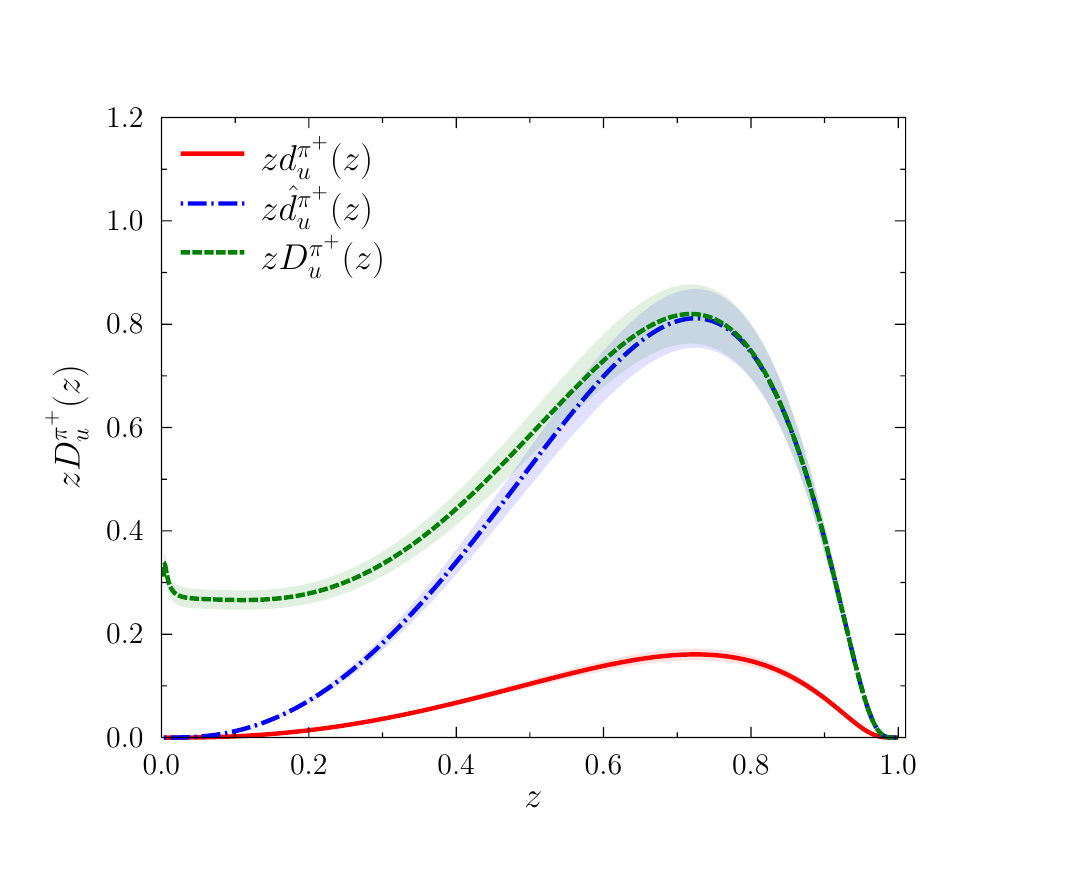} 
  \vspace*{-1.2cm}
\caption{The elementary fragmentation functions $d_u^{\pi^+}$ and $\hat d_u^{\pi^+}$ [normalized as in Eq.~\eqref{normsum}] and the pion fragmentation function 
$D_u^{\pi^+}$ at the scale $Q_0 =  0.63$\,GeV, where error bands are as in Fig.~\ref{Dfragfig}. }
\label{Dfragfull-elem} 
\end{figure}

\subsection{Quark-jet functions}

There is no reason to assume the quark fragments into a single pion described by the amplitude in Eq.~\eqref{EQ:fragLF}. On the contrary, one must consider the 
possibility that the fragmenting quark produces a cascade of mesons. In the  quark-jet model formulated by Field and Feynman~\cite{Field:1976ve}, the meson 
observed in a semi-inclusive process is one amongst many others which form a jet of hadrons. The probability for finding a pion with light-front momentum 
fraction $z$ in the jet is described by the full fragmentation or jet function $D_q^\pi(z)$, obtained via an iterative resummation of all possible fragmentations. 

With the convenient isospin decomposition~\cite{Ito:2009zc}, 
\begin{equation}
   D_q^\pi (z) \equiv \frac{1}{3} \, \big [ D^\pi_0 (z) + \tau_q \tau_\pi D^\pi_1 (z) \big ]\,  ,
 \label{FFisospindecomp}
\end{equation}
one can distinguish between  \emph{favored, unfavored\/} and \emph{neutral\/} fragmentation functions:
\begin{align}
   D_u^{\pi^+} \! & = \, D_d^{\pi^-}     \!  = \, D_{\bar{u}}^{\pi^-}   \! = \, D_{\bar{d}}^{\pi^+} \!  =  \,  \frac{1}{3} \, \big (D^\pi_0 + D^\pi_1 \big )  \, , \\
   D_u^{\pi^-}  \!  & = \, D_d^{\pi^+}   \!  = \, D_{\bar{u}}^{\pi^+}  \! = \, D_{\bar{d}}^{\pi^-}  \!  =  \,  \frac{1}{3}  \, \big (D^\pi_0 - D^\pi_1 \big )  \, , \\
   D_u^{\pi^0}  \!  & =\,  D_d^{\pi^0}  \!  = \, D_{\bar{u}}^{\pi^0}   \! = \, D_{\bar{d}}^{\pi^0} \!  =  \,  \frac{1}{3} \, D^\pi_0 \,  .
\end{align}
The unfavored functions denote those fragmentation functions where the hadronizing quark or antiquark \emph{is not} a valence quark of the pion. 
The functions $D^\pi_0 (z)$ and $D^\pi_1 (z)$ of Eq.~\eqref{FFisospindecomp} are obtained by solving the set of Volterra integral equations~\cite{Ito:2009zc},
\begin{align}
      \tfrac{2}{3}\, D_0^\pi (z)  & = \,  \hat d_q^\pi(z) +   \! \int_z^1  \hat d_q^\pi (1-z/ y ) \frac{D_0^\pi(y)}{y} \,  dy \, ,
    \label{D0int}  \\
     \tfrac{2}{3}\, D_1^\pi (z)  & = \,   \hat  d_q^\pi(z) - \tfrac{1}{3} \! \int_z^1 \hat   d_q^\pi  ( 1-z/ y ) \frac{D_1^\pi(y)}{y}  \,  dy\, ,
    \label{D1int}
\end{align}
which are normalized as $\int_0^1 z D_0^\pi (z) dz =1$ and  $\int_0^1  D_1^\pi (z) dz =3/4$. In the limit $z\to1$ the full fragmentation function~\eqref{FFisospindecomp} reduces to the 
elementary one, 
\begin{equation}
     D_u^\pi (z)  \xrightarrow{z\to 1}\,  \hat d_q^\pi(z)  \, ,
\end{equation}
since in this case the quark gives all its momentum to the initial pion, leaving no momentum left for a jet. The solution to Eqs.~\eqref{D0int} and \eqref{D1int}
is plotted in Fig.~\ref{Dfragfull-elem} for the pion fragmentation function $D_u^{\pi^+}$  along with the elementary fragmentation function. We observe 
that $D_u^{\pi^+}$ is overall enhanced and greater than elementary fragmentation function in the range $z\lesssim 0.6$.

\begin{figure}[t!] 
\hspace*{-2mm}
  \includegraphics[scale=0.47,angle=0]{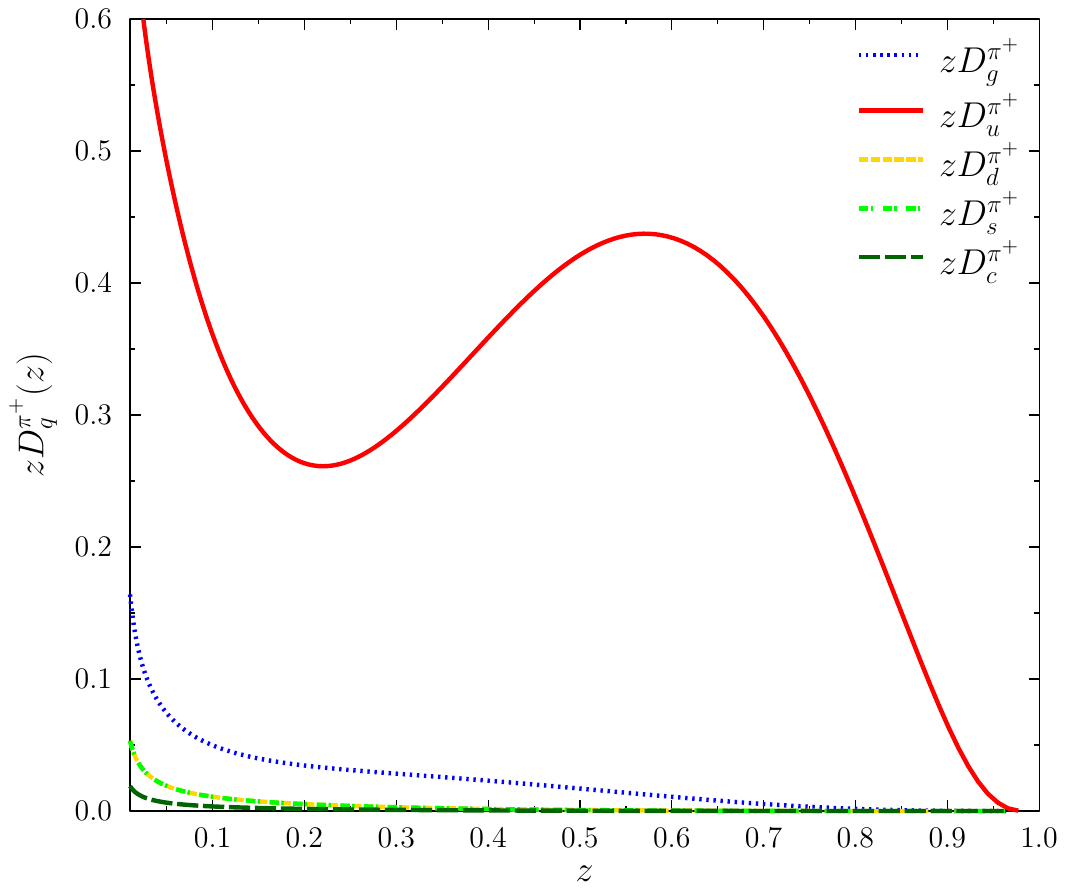} 
\caption{The $D_q^{\pi^+}$ fragmentation functions evolved from the model scale $Q_0$ to 2\;GeV and the evolution-generated gluon, $d$, $s$ and $c$ quark 
fragmentation functions. In the latter, the momentum fractions are of $d\bar d$,  $s\bar s$ and $c\bar c$ pairs generated by the initial $u$-quark in the 
DGLAP evolution~\cite{Hirai:2011si}.}
\label{Dfullevolved} 
\end{figure}

The moments of the fragmentation functions are defined as,
\begin{equation}
   \langle z \rangle_{D^\pi_q}^\mu  =  \int z D^\pi_q  (z,\mu) \, dz 
\end{equation}
and at the model scale $\mu=Q_0 = 0.63$\;GeV we find:
\begin{align}
  \langle z \rangle_{D^\pi_0}^{Q_0} = 0.91 \, , \quad  \langle z \rangle_{D^\pi_1 }^{Q_0}  =  0.51 \, , \quad  \langle z \rangle_{D_u^{\pi} }^{Q_0}  =  0.47\, .
\end{align}
After leading order DGLAP evolution of the fragmentation functions~\cite{Hirai:2011si} including $N_f=4$ thresholds, their moments at a scale $\mu=2$ GeV are:
\begin{subequations}
\begin{align}
 \langle z \rangle_{D_u^{\pi} }^{\mu} & =  0.31\, , \quad  \langle z \rangle_{D_{u,d\bar d}^{\pi} }^{\mu}   = 0.009 \, ,    \label{Duddbar}  \\ 
 \langle z \rangle_{D_{u,s\bar s}^{\pi}}^{\mu}  & =  0.009\, , \quad   \langle z \rangle_{D_{u,c\bar c}^{\pi} }^{\mu}  =  0.003  \, .
  \label{Dussbarccbar}
\end{align}
\end{subequations}
The moments in Eqs.~\eqref{Duddbar} and \eqref{Dussbarccbar} describe the momentum fractions of the $d\bar d$, $s\bar s$ and $c\bar c$ pairs generated 
in the evolution equations from the original $u$-quark. The corresponding pion fragmentation functions evolved to the scale $\mu = 2$ GeV are drawn as functions 
of $z$ in Fig.~\ref{Dfullevolved}. Notably, as seen from comparison of  Figs.~\ref{Dfragfull-elem} and \ref{Dfullevolved}, the fragmentation function $D_u^{\pi^+}$ 
is strongly enhanced at small $z$, while the hump structure persists even though the original maximum is considerably lower. In the evolution we assume that the pion 
is made of $u$ and $\bar d$ quarks only at the model scale. It has been advocated~\cite{Xing:2023pms} that a singlet jet equation in QCD should involve gluon 
contributions to the cascade because of gluon and $q\bar q$ mixing, which implies that $D_g^\pi(z,Q_0) \neq 0$ contributes to momentum conservation. 
Doing so would obviously lead to an enhancement of $D_g^{\pi^+} (z,\mu)$ in Fig.~\ref{Dfullevolved}, that is  if we assumed some model for that fragmentation function 
at our model scale $Q_0$.


\section{Conclusive Remarks}
\label{sec4}

We present the first calculation of the elementary $q\to \pi$ fragmentation function in the DSE-BSE framework employing a cut-diagram representation of
the fragmentation process and a convenient algebraic representations of the pion's BSA and quark propagators. The latter allows us to analytically integrate 
over the light-front variables $k^+$ and $k^-$ while the transverse momentum can be integrated numerically. The fragmentation function we obtain is in qualitative
agreement with an earlier calculation in the NJL model, but is about 10--12\% enhanced for momentum fractions $z \lesssim 0.8$.

This elementary fragmentation function feeds into the kernel of a quark jet fragmentation equation, which is solved to obtain the full pion fragmentation function.
The resulting functions $D^{\pi^+}_u$ and $D^{\pi^-}_u$, amongst others, describe the probability that a $u$ quark escaping the collision region produces
a positively or negatively charged pion, respectively, thereby transferring a fraction of its light-front momentum $z$. 

The present approach is readily applicable to other light and heavy meson fragmentation functions and calculations are underway. As a light
quark can fragment in a cascade of mesons (and baryons) other than pions, this leads to a set of coupled quark jet fragmentation functions that must be solved
self-consistently~\cite{Casey:2012ux}. Future improvements must also consider gluon contributions to the isospin-singlet fragmentation function $D^\pi_0$ at 
the initial hadronic scale, as one expects gluon and $q\bar q$ mixing. The fragmentation of a gluon poses a more daunting challenge, as the topology of their 
hadronization involves higher-order diagrams and a robust ansatz for the quark-gluon vertex. Nevertheless, their calculation will reveal unknown aspects of 
hadronization and shed light on the underlying confinement mechanism.


\acknowledgements 

We are indebted to Ian Clo\"et for very helpful discussions. B.\,E. acknowledges support by the S\~ao Paulo Research Foundation (FAPESP), grant no.~2023/00195-8 and the 
National Council for Scientific and Technological Development (CNPq), grant no.~409032/2023-9, and  R.\,C.\,S is supported by a postdoctoral CAPES-PIPD fellowship.
B.\,E. and R.\,C.\,S are members of the Brazilian network project \emph{INCT-F\'isica Nuclear e Aplica\c{c}\~oes\/}, grant no.~464898/2014-5.


\appendix

\section{Algebraic representation of propagators and Bethe-Salpeter amplitudes}
\label{appendix1}

Since we work with light-front variables, a suitable algebraic representation for the propagators and bound-state amplitudes that enter the fragmentation 
function~\eqref{EQ:fragLF} is required. A sum of complex conjugate mass poles~\cite{Bhagwat:2002tx,El-Bennich:2016qmb} provides an adequate reproduction of the numerical 
solution of the scalar and vector dressing  functions  of the quark propagator $S(k) =  -i \sigma_V(k^2)\gamma \cdot k + \sigma_S(k^2)$,
\begin{align}
  \sigma_S (k^2)  & = \sum_{n=1}^2 \left [ \frac{z_n m_n}{k^2+m_n^2}+\frac{z_n^* m_n^*}{k^2+ {m^*_n}^2} \right ] , \\
  \sigma_V (k^2)  & = \sum_{n=1}^2 \left [ \frac{z_n}{k^2+m_n^2}+\frac{z_n^*}{k^2 + {m^*_n}^2}   \right ]  ,  \\
\end{align}
where $m_n$ are complex-valued light-quark mass scales and $z_n$ are complex coefficients  fitted to the numerical values of $ \sigma_S(k^2)$ and $\sigma_V  (k^2)$.
Their values are found in Table~2 of Ref.~\cite{Serna:2020txe}.

The Lorentz-invariant amplitudes of the BSA~\eqref{PS-BSA} are expanded in terms of Chebyshev moments,  $\mathcal{F}_{im} ( k, P )$,
\begin{equation}
   \mathcal{F}_i ( k, P,z_k ) = \sum_{m=0}^{\infty} \mathcal{F}_{im} ( k, P )\, U_m (z_k) \ ,
\label{chebyshev}   
\end{equation}
which results in a faster convergence in solving the BSE. We consider $m=3$ Chebyshev polynomials $U_m (z_k )$ of the second kind, where $z_k=k \cdot P /|k \| P|$ 
is an angle between $k$ and $P$. The eigenvalue problem for the  vector  $\boldsymbol{\mathcal{F}} := \{ \mathcal{F}_1,  \mathcal{F}_2,  \mathcal{F}_3, \mathcal{F}_4 \} 
= \{ E_\pi, F_\pi, G_\pi, H_\pi \}$ that describes the pseudoscalar meson is solved by means of Arnoldi factorization implemented with the \texttt{ARPACK} 
library~\cite{Lehoucq1998}. For details we refer, for instance, to Refs.~\cite{Blank:2010bp,Rojas:2014aka}.

One can separate the BSA in even and odd components,
\begin{equation}
\label{BSA-kaon-D}
   \mathcal{F}_i (k,P) = \mathcal{ F}^0_i (k,P) + k\cdot P\, \mathcal{F}^1_i (k,P) \, ,  
\end{equation}
where $\mathcal{F}^{0,1}_i (k,P)$ are even under $k\cdot P \rightarrow   -k\cdot P$ and  $\mathcal{F}^1_i (k,P) \equiv 0$ for flavorless pseudoscalar mesons, 
as they are eigenstates of the charge-conjugation operator defined as,
\begin{equation}
   \Gamma_M (k,P) \, \stackrel{C}{\longrightarrow} \  \bar{\Gamma}_M (k,P) :=C\, \Gamma^{T}_M (-k,P) C^{T} \ .
\end{equation}
In here, we assume $SU(2)$ isospin symmetry with $m_u=m_d$. The constraint that the covariant basis~\eqref{PS-BSA} must satisfy $\bar \Gamma_M (k,P)  
= \lambda_c \Gamma_M (k,P)$ with $\lambda_c  = +1$ for pseudoscalar mesons is the reason that $\mathcal{F}_i^0 (k,P)$ is even.

The four scalar amplitudes $\mathcal{F}_i (k,P) = \mathcal{ F}^0_i (k,P)$ are parametrized with a generalized Nakanishi integral representation~\cite{Nakanishi:1965zza,
Nakanishi:1965zz},
\begin{equation}
  \mathcal{F}_i (k,P) = \sum_{j=1}^{N} \int_{-1}^{1} \! d\alpha\,  \rho_j (\alpha)\, \frac{U_j \Lambda^{2n_j}}{\Delta^{n_j} (k,\alpha,\Lambda) }    \, ,
 \label{NakanishiBSA} 
\end{equation}
where $\Delta=k^{2} + \alpha\, k\cdot P+\Lambda^{2}$, $N=3$ and the spectral density $\rho_j (\alpha)$ is given by,
\begin{equation}
   \rho_j  (\alpha) = \tfrac{1}{2} \! \left ( C^{\nicefrac{1}{2}}_{0}(\alpha)+\sigma_j C^{\nicefrac{1}{2}}_{2}(\alpha)  \right ) \, .
 \label{rhropion}   
\end{equation}
$C^{\nicefrac{1}{2}}_{0}(\alpha)$ and $C^{\nicefrac{1}{2}}_{2}(\alpha)$ are Gegenbauer polynomials of order $1/2$. The parameters $U_j$, $\Lambda$, $n_j$ and $\sigma_j$ 
for the pion are listed in Table~IV of Ref.~\cite{daSilveira:2022pte}. The amplitudes $\mathcal{F}_i (k,P)$  are fitted to the sum of the 0th and 2nd Chebyshev moments.



\end{document}